\documentclass[10pt]{article}
\usepackage[OE]{express}
\usepackage{graphicx}
\usepackage{amsmath,amssymb,bm}
\usepackage{subfigure}
\usepackage{siunitx}
\usepackage{dsfont}

\begin{document}

\title{Surface modes in plasmonic Bragg fibers with negative average
permittivity}
\author{Hanying Deng,\authormark{1} Yihang Chen,\authormark{1,7} Nicolae C. Panoiu,\authormark{2} Boris A. Malomed,\authormark{3,4} and Fangwei Ye \authormark{5,6,8} }

\address{\authormark{1}School of Physics and Telecommunication Engineering, South
China Normal University, Guangzhou 510006, China\\
\authormark{2}Department of Electronic and Electrical Engineering, University College London,
Torrington Place, London WC1E7JE, United Kingdom\\

\authormark{3}Department of Physical Electronics, School of
Electrical Engineering, Faculty of Engineering, Tel Aviv University,
Tel Aviv 69978, Israel\\
\authormark{4}ITMO University, St. Petersburg 197101, Russia\\

\authormark{5}School of Physics and Astronomy, Shanghai Jiao Tong University, Shanghai
200240, China\\
\authormark{6}{Department of Physics, Zhejiang Normal University, Jinhua 321004, China}

\authormark{7}yhchen@scnu.edu.cn\\
\authormark{8}fangweiye@sjtu.edu.cn}

\begin{abstract}
We investigate surface modes in plasmonic Bragg fibers composed of
nanostructured coaxial cylindrical metal-dielectric multilayers. We
demonstrate that the existence of surface modes is determined by the sign of
the spatially averaged permittivity of the plasmonic Bragg fiber, $\bar{%
\varepsilon}$. Specifically, localized surface modes occur at the interface
between the cylindrical core with $\bar{\varepsilon}<0$ and the outermost
uniform dielectric medium, which is similar to the topologically protected
plasmonic surface modes at the interface between two different
one-dimensional planar metal-dielectric lattices with opposite signs of the
averaged permittivity. Moreover, when increasing the number of
dielectric-metal rings, the propagation constant of surface modes with
different azimuthal mode numbers is approaching that of surface plasmon
polaritons formed at the corresponding planar metal/dielectric interface.
Robustness of such surface modes of plasmonic Bragg fibers is demonstrated
too.
\end{abstract}








\section{Introduction}

Surface modes induced by nontrivial topological mechanisms have recently
drawn much attention in optics. Thanks to the topological protection,
topological surface modes are intrinsically robust against structural
perturbations \cite{Poliee,Cheng,llnp,Wbing}. A variety of optical systems
supporting such surface modes have been proposed and demonstrated. The
simplest example is analogous to the celebrated Su-Schrieffer-Heeger (SSH)
model for polyacetylene \cite{su1979solitons}, in which a chain of sites
with alternating coupling constants exhibits two topologically distinct
phases, and topologically protected interfacial modes exist at their
interface. The photonic realization of the SSH model was demonstrated in
dimerized dielectric waveguides \cite{henningol}, dielectric nanoparticles
\cite{pprlalobor}, and metallic nanodisks \cite{nanoscale}, as well as in
graphene plasmonic waveguide arrays \cite{ge2015topological}. These
structures, which implement the SSH model in photonics, are discrete one, as
concerns the arrangement of their optical elements.

%
%
%
%

Topological surface modes can also be realized in one-dimensional (1D)
continuous periodic systems \cite{zchenol,prxctchan,Deng}. In particular,
the Zak phase, which is a special kind of the Berry phase defined for
photonic bands of 1D systems, characterizes topological properties of such
periodic systems \cite{zakberry}. Interestingly, it was found that the Zak
phase of plasmonic superlattices, composed of alternating metallic and
dielectric layers, is determined by the sign of the spatially averaged value
of their permittivity \cite{Deng}. Topologically protected plasmonic surface
modes exist at the interface between two plasmonic lattices with opposite
signs of the average permittivity, and these modes may be regarded as a
generalization of conventional surface plasmonic polaritons (SPPs) occurring
at the interface between dielectric and metallic materials. This means that
for 1D plasmonic periodic systems the spatial average of the permittivity,
which has a more intuitive physical meaning, acts as an alternative to the
Zak phase in characterizing topological properties of the structure --
namely, in defining the existence of topological surface modes. This raises
an important question: can the sign of the average permittivity also
determine the existence of plasmonic surface modes in metallic-dielectric
structures, beyond the 1D case, such as in plasmonic Bragg fibers composed
of coaxial cylindrical metal-dielectric multilayers?

In this paper, we consider a plasmonic Bragg fiber composed of coaxial
cylindrical dielectric-metal multilayers and investigate physical properties
of localized surface modes in such fibers by performing the mode analysis
and direct beam-propagation simulations. We find that, similarly to the case
of topological surface modes in planar metal-dielectric multilayers, surface
modes exist at the interface between the core of the plasmonic Bragg fiber
with a negative average permittivity and the outermost uniform dielectric
medium. We also find that, with the increase of number of dielectric-metal
rings of the plasmonic Bragg fiber, the propagation constant of all the
modes, including the fundamental and higher-order ones, approaches that of
surface plasmon polaritons formed at the planar interface between the metal
and dielectric media. Finally, using direct numerical simulations, the
surface modes of plasmonic Bragg fibers are found to be robust against
structural disorder. The paper is organized as follows. In the next section
we introduce the theoretical model which is used to describe optical
properties of our structure. Then, in Sec.~\ref{sec:disc}, we present and
discuss the main results of our study. Finally, we summarize the results in
Sec.~\ref{sec:concl}.

\section{The theoretical model and transfer-matrix formalism}
\label{sec:model} The plasmonic Bragg fiber considered in this study is built as a series of
coaxial metal-dielectric cylindrical shells, as schematically shown in Fig. 1. The electromagnetic
field is assumed to propagate along the common axis of the cylindrical shells, $z$. To make the
analysis more specific, we assume that the metallic and dielectric layers are made of silver and
silicon, respectively. The complex permittivity of
the metal (silver) can be calculated by using the Drude model \cite%
{johnsonoptical}, $\varepsilon _{M}=1-\omega _{p}^{2}/[\omega (\omega +i\nu
)]$, with plasma and damping frequencies being $\omega _{p}=%
\SI{13.7e15}{\radian\per\second}$ and $\nu =\SI{2.7e14}{\radian\per\second}$%
, respectively. Note that the effect of interband transitions can be added
to the Drude permittivity without qualitatively changing main conclusions of
our analysis. The permittivity of the dielectric (silicon) is $\varepsilon
_{D}=12.25$. The thickness of the metal layer is fixed to $t_{M}=%
\SI{25}{\nano\meter}$.

In order to be possible to use the operating wavelength to conveniently tune the
average permittivity of the\ plasmonic Bragg fiber, $\bar{\varepsilon}$, from being negative,
passing through zero, to positive values, we assume that, at $\lambda =\SI{1550}{\nano\meter}$
($\varepsilon _{M}=-125.39$), the average permittivity of each metal-dielectric ring pair is tuned
to zero; this of course also means that the average permittivity of the whole Bragg structure is
zero.  As we are concerned with TM-polarized waves and $\lambda \gg t_M^n, t_D^n$, this leads to
the following expression at the corresponding wavelength:
\begin{equation}
\bar{\varepsilon}_{n}=\frac{\varepsilon _{D}S_{D}^{n}+\varepsilon
_{M}S_{M}^{n}}{S_{D}^{n}+S_{M}^{n}}=0,~~~~~n=1,2,3,\cdots
\end{equation}%
where $S_{M}^{n}=\pi \left[ (r_{M}^{n})^{2}-(r_{M}^{n}-t_{M})^{2}\right] $
and $S_{D}^{n}=\pi \left[ (r_{M}^{n}+t_{D}^{n})^{2}-(r_{M}^{n})^{2}\right] $
are the cross-sectional areas of the $n$th metallic and dielectric
cylindrical shells, respectively.  $r_{M}^{n}$ and $t_{D}^{n}$ are the inner radius of the
$n$th metallic cylindrical shell and the thickness of the $n$th dielectric cylindrical shell,
respectively. As the thickness of the metallic shell is assumed to be fixed, thus the values
for the series of the dielectric shells, $t_{D}^{n}$, can be obtained from Eq. (1).
\begin{figure}[t]
\centering\includegraphics[width=12cm]{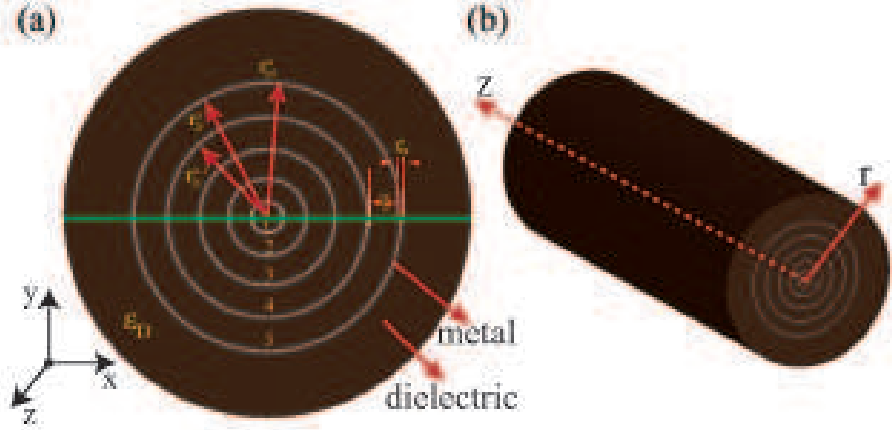}
\caption{(a) The cross-section of a plasmonic Bragg fiber composed of
alternating coaxial metal-dielectric cylindrical shells. (b) A schematic
structure of the plasmonic Bragg fiber.}
\end{figure}

Thus, thanks to the strong dispersion of the metallic permittivity,  the average permittivity of
such a plasmonic Bragg fiber,
\begin{equation}
\bar{\varepsilon}=\frac{\sum_{n}(\varepsilon _{D}S_{D}^{n}+\varepsilon
_{M}S_{M}^{n})}{\sum_{n}(S_{D}^{n}+S_{M}^{n})}=\frac{\sum_{n}\bar{\varepsilon%
}_{n}(S_{D}^{n}+S_{M}^{n})}{\sum_{n}(S_{D}^{n}+S_{M}^{n})},
\end{equation}%
can be shifted from negative to positive values by simply changing the operational wavelength. In
our structure, the outermost layer is a homogeneous dielectric (silicon), whose thickness does not
have to satisfy $\bar{\varepsilon}_{n}=0$. Since the imaginary part of the permittivity of the
metal is very small as compared to its real part($\epsilon_M=-125.39-2.84 i$), the influence of
the metal loss on the results are found to be negligible. Nevertheless, in the following analysis
we have taken into account this small imaginary part, unless otherwise stated.

The electromagnetic field in the plasmonic Bragg fiber can be calculated by
employing the transfer-matrix method. Thus, the $z$-component of the
electric and magnetic fields in the $n$th dielectric layer, which are a
solution of the Maxwell equations expressed in cylindrical coordinates ($r$,
$\theta $, $z$), can be written in the following form \cite{Yen}:
\begin{equation}
\begin{array}{l}
{E_{z}}(r,\theta ,z)=[A_{D}^{n}{I_{m}}({k_{D}}r)+B_{D}^{n}{K_{m}}({k_{D}}%
r)]\cos (m\theta ){e^{i{\beta _{m}}z}}, \\
{H_{z}}(r,\theta ,z)=[C_{D}^{n}{I_{m}}({k_{D}}r)+D_{D}^{n}{K_{m}}({k_{D}}%
r)]\sin (m\theta ){e^{i{\beta _{m}}z}},%
\end{array}%
\end{equation}%
where ${k_{D}}={\left[ \beta _{m}^{2}-(\omega /c)^{2}\varepsilon _{D}\right]
}^{1/2}$ and, in the $n$th metallic layer, the fields are
\begin{equation}
\begin{array}{l}
{E_{z}}(r,\theta ,z)=[A_{M}^{n}{I_{m}}({k_{M}}r)+B_{M}^{n}{K_{m}}({k_{M}}%
r)]\cos (m\theta ){e^{i{\beta _{m}}z}}, \\
{H_{z}}(r,\theta ,z)=[C_{M}^{n}{I_{m}}({k_{M}}r)+D_{M}^{n}{K_{m}}({k_{M}}%
r)]\sin (m\theta ){e^{i{\beta _{m}}z}},%
\end{array}%
\end{equation}%
where ${k_{M}}\equiv {\left[ \beta _{m}^{2}-(\omega /c)^{2}\varepsilon _{M}%
\right] }^{1/2}$, $I_{m}$ and $K_{m}$ are the modified Bessel functions of
the first and the second kind, respectively, $\beta _{m}$ is the propagation
constant, $m$ is the azimuthal mode number which defines the order of the
mode, $\omega $ is the angular frequency, and $c$ is the speed of light in
vacuum. Moreover, $A_{M}^{n}$, $B_{M}^{n}$, $C_{M}^{n}$, and $D_{M}^{n}$ ($%
A_{D}^{n}$, $B_{D}^{n}$, $C_{D}^{n}$, and $D_{D}^{n}$) are modal amplitude
coefficients within the $n$th metallic (dielectric) layers.

The transverse components of the fields can be expressed in terms of $E_z$
and $H_z$ using the following relations:
\begin{eqnarray}
{E_r} = \frac{{i{\beta _m}}}{{{\omega ^2}\mu \varepsilon - \beta _m^2}}\left(%
\frac{{\partial {E_z}}}{{\partial r}} + \frac{{\omega \mu }}{{\beta _m}}%
\frac{{\partial {H_z}}}{{r\partial \theta}}\right), \\
{E_\theta } = \frac{{i{\beta _m}}}{{{\omega ^2}\mu \varepsilon - \beta _m^2}}%
\left(\frac{{\partial {E_z}}}{{r\partial \theta }} - \frac{{\omega \mu }}{{%
\beta _m}}\frac{{\partial {H_z}}}{{\partial r}}\right), \\
{H_r} = \frac{{i{\beta _m}}}{{{\omega ^2}\mu \varepsilon - \beta _m^2}}\left(%
\frac{{\partial {H_z}}}{{\partial r}} - \frac{{\omega \varepsilon }}{{\beta
_m}}\frac{{\partial {E_z}}}{{r\partial \theta}}\right), \\
{H_\theta } = \frac{{i{\beta _m}}}{{{\omega ^2}\mu \varepsilon - \beta _m^2}}%
\left(\frac{{\partial {H_z}}}{{r\partial \theta }} + \frac{{\omega
\varepsilon }}{{\beta _m}}\frac{{\partial {E_z}}}{{\partial r}}\right),
\end{eqnarray}
where $\mu=\mu_{0}$ is the magnetic permeability of vacuum.

Once the electromagnetic field in the $n$th dielectric layer is known, one
can easily find the field in the $n$th metallic layer by applying the
boundary conditions at the interface between the $n$th dielectric and
metallic layers. The continuity of the tangential field components $E_{z}$, $%
H_{z}$, $E_{\theta }$, and $H_{\theta }$ yields:
\begin{equation}
M_{M}(\rho )\left(
\begin{array}{l}
A_{M}^{n} \\
B_{M}^{n} \\
C_{M}^{n} \\
D_{M}^{n}%
\end{array}%
\right) =M_{D}(\rho )\left(
\begin{array}{l}
A_{D}^{n} \\
B_{D}^{n} \\
C_{D}^{n} \\
D_{D}^{n}%
\end{array}%
\right) ,
\end{equation}%
with matrices
\begin{equation}
  {M_M}(\rho ) = \left( {\begin{array}{*{20}{c}}
{{I_m}(\varrho _{M} )}&{{K_m}(\varrho _{M})}&0&0\\
{\frac{{\omega {\varepsilon _0}{\varepsilon _M}}}{{\varrho _{M} }}{{I'}_m}(\varrho _{M} )}&{\frac{{\omega {\varepsilon _0}{\varepsilon _M}}}{{\varrho _{M} }}{{K'}_m}(\varrho _{M} )}&{\frac{m\rho}{{{{{\varrho _{M}^2}}} }}{I_m}(\varrho _{M} )}&{\frac{m\rho}{{{{\varrho _{M}^2}} }}{K_m}(\varrho _{M} )}\\
0&0&{{I_m}(\varrho _{M} )}&{{K_m}(\varrho _{M} )}\\
{\frac{m\rho }{{{{{\varrho _{M}^2}}}}}{I_m}(\varrho _{M} )}&{\frac{m\rho }{{{{{\varrho _{M}^2}}} }}{K_m}(\varrho _{M} )}&{\frac{{\omega {\mu _0}{\mu _M}}}{{\varrho _{M} }}{{I'}_m}(\varrho _{M} )}&{\frac{{\omega {\mu _0}{\mu _M}}}{{\varrho _{M} }}{{K'}_m}(\varrho _{M} )}
\end{array}} \right)
\end{equation}
and
\begin{equation}
  {M_D}(\rho ) = \left( {\begin{array}{*{20}{c}}
{{I_m}(\varrho _{D} )}&{{K_m}(\varrho _{D})}&0&0\\
{\frac{{\omega {\varepsilon _0}{\varepsilon _D}}}{{\varrho _{D} }}{{I'}_m}(\varrho _{D} )}&{\frac{{\omega {\varepsilon _0}{\varepsilon _D}}}{{\varrho _{D} }}{{K'}_m}(\varrho _{D} )}&{\frac{m\rho}{{{{{\varrho _{D}^2}}} }}{I_m}(\varrho _{D} )}&{\frac{m\rho}{{{{\varrho _{D}^2}} }}{K_m}(\varrho _{D} )}\\
0&0&{{I_m}(\varrho _{D} )}&{{K_m}(\varrho _{D} )}\\
{\frac{m\rho }{{{{{\varrho _{D}^2}}}}}{I_m}(\varrho _{D} )}&{\frac{m\rho }{{{{{\varrho _{D}^2}}} }}{K_m}(\varrho _{D} )}&{\frac{{\omega {\mu _0}{\mu _D}}}{{\varrho _{D} }}{{I'}_m}(\varrho _{D} )}&{\frac{{\omega {\mu _0}{\mu _D}}}{{\varrho _{D} }}{{K'}_m}(\varrho _{D} )}
\end{array}} \right)
\end{equation}
Here, $\rho =r_{D}^{n}$ is the inner radius of the $n$th dielectric cylindrical shell, and
$\varrho _{M,D}=k_{M,D}\rho $. Moreover, Eq. (9) can be rewritten as:
\begin{equation}
\left(
\begin{array}{l}
A_{M}^{n} \\
B_{M}^{n} \\
C_{M}^{n} \\
D_{M}^{n}%
\end{array}%
\right) =M_{M}^{-1}(\rho ){M_{D}}(\rho )\left(
\begin{array}{l}
A_{D}^{n} \\
B_{D}^{n} \\
C_{D}^{n} \\
D_{D}^{n}%
\end{array}%
\right) .
\end{equation}

Similarly, knowing coefficients of the electromagnetic fields in the $n$th
metallic layer, one can determine their counterparts in the $(n+1)$th
dielectric layer by using the following matrix relation:
\begin{equation}
\left(
\begin{array}{l}
A_{D}^{n+1} \\
B_{D}^{n+1} \\
C_{D}^{n+1} \\
D_{D}^{n+1}%
\end{array}%
\right) =M_{D}^{-1}(\rho ^{\prime }){M_{M}}(\rho ^{\prime })\left(
\begin{array}{l}
A_{M}^{n} \\
B_{M}^{n} \\
C_{M}^{n} \\
D_{M}^{n}%
\end{array}%
\right) ,
\end{equation}%
where $\rho ^{\prime }=r_{M}^{n}$ is the outer radius of the $n$th metallic
cylindrical shell.

Equations (12) and (13) can be used iteratively to relate the amplitude
coefficients in the first dielectric layer, i.e., $A_{D}^{1}$, $B_{D}^{1}$, $%
C_{D}^{1}$, and $D_{D}^{1}$, to their counterparts in the outermost
dielectric layer, i.e., $A_{D}^{n+1}$, $B_{D}^{n+1}$, $C_{D}^{n+1}$, and $%
D_{D}^{n+1}$. The final result of this operation can be written as:
\begin{equation}
\left( \begin{array}{l}
A_D^{n+1}\\
B_D^{n+1}\\
C_D^{n+1}\\
D_D^{n+1}
\end{array} \right) = M_D^{ - 1}(r_M^n){M_M}(r_M^n) \cdots M_M^{ - 1}(r_D^1){M_D}(r_D^1)\left( \begin{array}{l}
A_D^1\\
B_D^1\\
C_D^1\\
D_D^1
\end{array} \right) = \left( {\begin{array}{*{20}{c}}
{{t_{11}}}&{{t_{12}}}&{{t_{13}}}&{{t_{14}}}\\
{{t_{21}}}&{{t_{22}}}&{{t_{23}}}&{{t_{24}}}\\
{{t_{31}}}&{{t_{32}}}&{{t_{33}}}&{{t_{34}}}\\
{{t_{41}}}&{{t_{42}}}&{{t_{43}}}&{{t_{44}}}
\end{array}} \right)\left( \begin{array}{l}
A_D^1\\
B_D^1\\
C_D^1\\
D_D^1
\end{array} \right).
\end{equation}

Two additional boundary conditions should be considered to determine the
electromagnetic field in the plasmonic Bragg fiber. First, the fields in the
first metallic layer must be finite, while $K_{m}$ is has a singularity at $%
r=0$. This requires that $B_{D}^{1}=D_{D}^{1}=0$. Second, the amplitude of
the electromagnetic waves in the outermost dielectric layer (the cover)
should be finite. This requires that $A_{D}^{n+1}=C_{D}^{n+1}=0$, since $%
I_{m}$ is infinite at $r\rightarrow \infty $. Thus Eq. (14) can be written
as:
\begin{equation}
\begin{array}{l}
A_{D}^{n+1}={t_{11}}A_{D}^{1}+{t_{13}}C_{D}^{1}=0, \\
C_{D}^{n+1}={t_{31}}A_{D}^{1}+{t_{33}}C_{D}^{1}=0,%
\end{array}%
\end{equation}%
which, in turn, may be expressed as
\begin{equation}
\left( {\begin{array}{*{20}{c}}
{{t_{11}}}&{{t_{13}}}\\
{{t_{31}}}&{{t_{33}}}
\end{array}} \right)\left( \begin{array}{l}
A_D^1\\
C_D^1
\end{array} \right) = 0.
\end{equation}
In order for Eq. (16) to have non-trivial solutions, the determinant of the matrix must be zero,
which gives:
\begin{equation}
\frac{t_{11}}{t_{31}} = \frac{t_{13}}{t_{33}}.
\end{equation}

Once the structure of the plasmonic Bragg fiber structure is chosen and the
frequency is fixed, Eq. (17) only depends on ${\beta _{m}}$. Therefore, the
solution of Eq. (17) gives us the propagation constant of any optical mode
of the fiber. After finding the value of ${\beta _{m}}$, one can determine
values of $A_{D}^{1}$ and $C_{D}^{1}$ from Eq. (16). Subsequently, by
combining this result with Eqs. (12) and (13), one can calculate the
electromagnetic field in any layer of the plasmonic Bragg fiber.

\begin{figure}[th]
\centering\includegraphics[width=12cm]{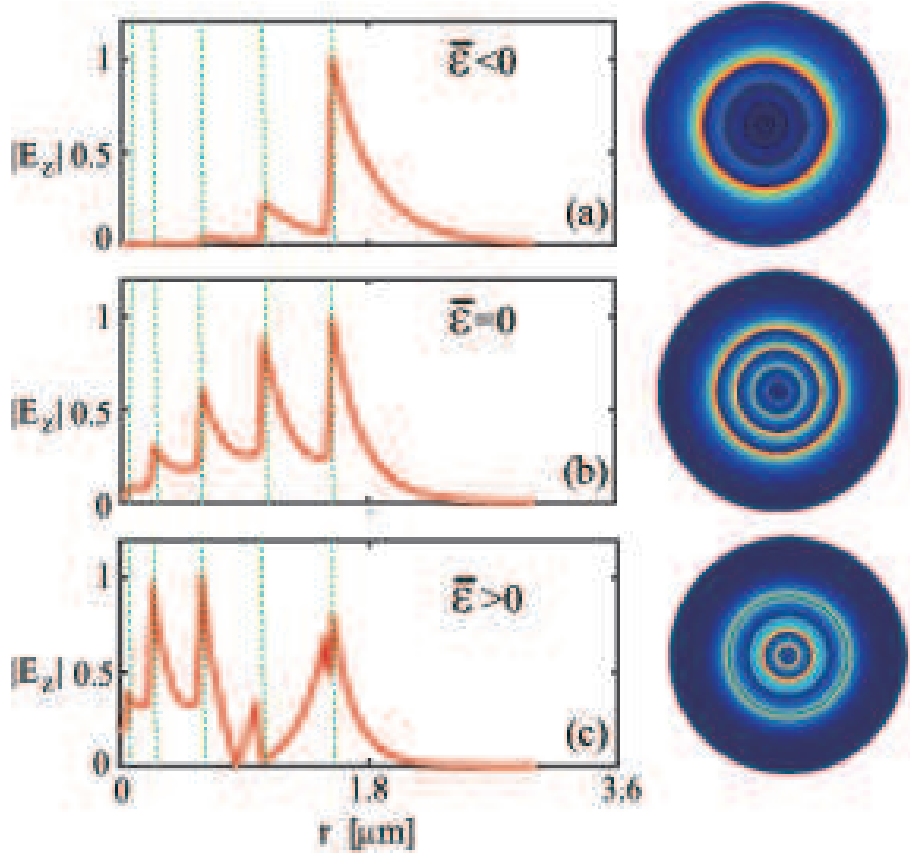}
\caption{The left (right) panel shows $|E_{z}|$ profiles of the fundamental
mode of the plamonic Bragg fiber with (a) $\bar{\protect\varepsilon}<0$, (b)
$\bar{\protect\varepsilon}=0$, and (c) $\bar{\protect\varepsilon}>0$,
obtained by using the transfer-matrix method (COMSOL simulations). In all
panels, the plasmonic Bragg fiber consists of five metal-dielectric ring
units with $t_{M}=\SI{25}{\nano\meter}$ and $\protect\varepsilon _{D}=12.25$%
. The average permittivities corresponding to the operational wavelength
are: (a) $\bar{\protect\varepsilon}=-11.4+0.72i$ at $\protect\lambda =
\SI{2.2}{\micro\meter}$, $\protect\varepsilon _{M}=-253.5+8.124i$; (b) $\bar{%
\protect\varepsilon}=0$ at $\protect\lambda =\SI{1.55}{\micro\meter}$, $%
\protect\varepsilon _{M}=-125.39+2.84i$; (c) $\bar{\protect\varepsilon}%
=2.2+0.1825i$ at $\protect\lambda =\SI{1.39}{\micro\meter}$, $\protect%
\varepsilon _{M}=-100.66+2.05i$. The green dashed lines indicate the interfaces
between the metallic and dielectric layers. }
\end{figure}

\begin{figure}[th]
\centering\includegraphics[width=12cm]{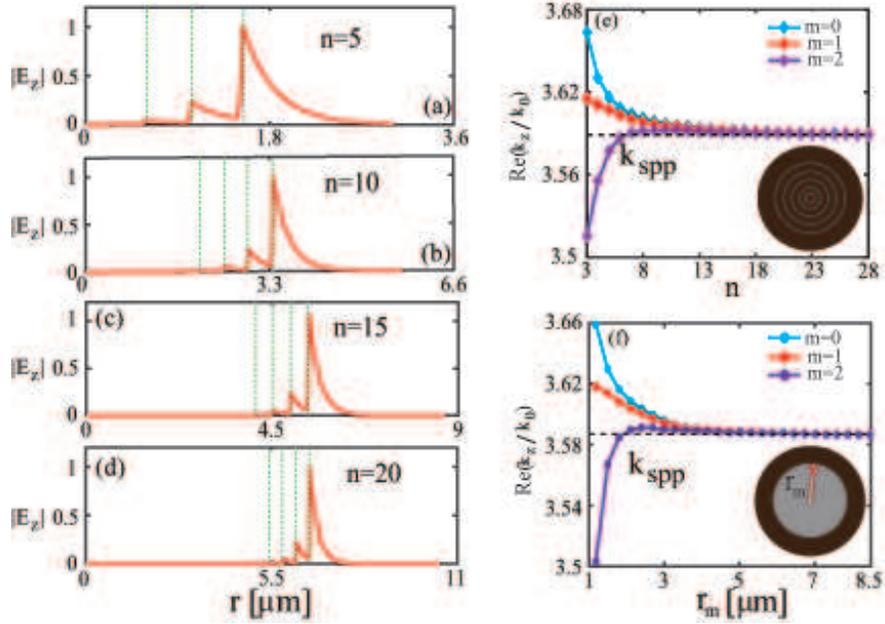}
\caption{(a), (b), (c), (d) Profiles of the electric field, $|E_{z}|$, of
the surface modes located at the interface between a plasmonic Bragg fiber
with $\bar{\protect\varepsilon}<0$ and a uniform dielectric medium with $%
\protect\varepsilon _{D}=12.25$. From top to bottom panels, the number of
metal-dielectric rings, $n$, of the plasmonic Bragg fiber increases. (e) The
dependence of the effective mode index of the plasmonic surface modes with $%
m=0,1,2$, on the the number of the metal-dielectric rings. The black dashed
line denotes the propagation constant of the SPPs at the corresponding
planar metallic-dielectric interface. (f) The same as in (e) but with the
plasmonic Bragg fiber replaced by a homogeneous metallic cylinder with the
same radius. The green dashed lines indicate the interfaces
between the metallic and dielectric layers.}
\end{figure}

\begin{figure}[th]
\centering\includegraphics[width=12cm]{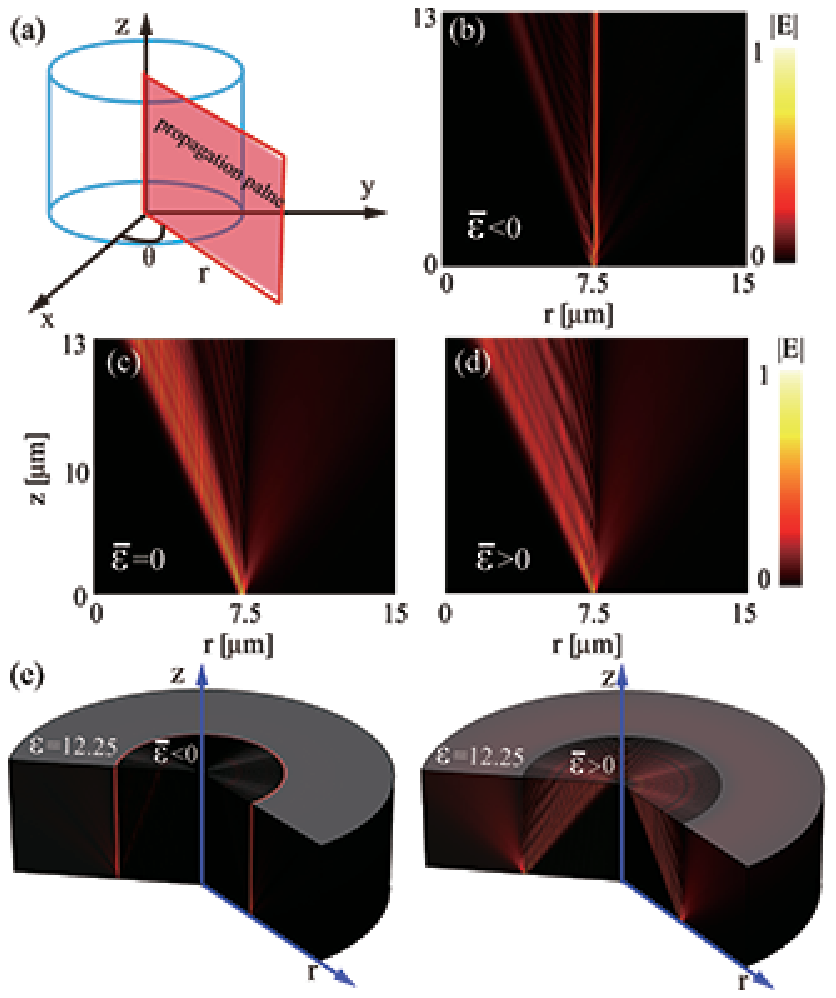}
\caption{Simulation of beam propagation in plasmonic Bragg fibers. The beam
dynamics is shown in the radial plane, as indicated in (a). A TM-polarized
Gaussian beam is injected normally at the interface between a homogeneous
medium with $\protect\varepsilon _{D}=12.25$ and a plasmonic Bragg fiber
with negative (b), vanishing (c), and positive (d) average permittivity. The
plasmonic Bragg fiber is composed of $n=30$ dielectric-metal rings.  (e) A 3D view of
beam propagation corresponding to (b) and (d), respectively.  }
\end{figure}

\section{Numerical results and discussion}

\label{sec:disc} We now employ the transfer-matrix formalism to calculate the field distribution
in the plasmonic Bragg fibers and investigate the existence of localized surface states. First, we
consider a plasmonic Bragg fiber composed of five dielectric-metal rings and surrounded by a
homogeneous dielectric cladding with permittivity $\varepsilon _{D}=12.25$, as per Fig. 1. As
discussed in the preceding section, the spatially averaged permittivity of such a plasmonic Bragg
fiber can be tuned simply by varying the operational wavelength. Figure 2 shows the electric-field
distribution of the fundamental mode, namely the mode with $m=0$, of the plasmonic Bragg fiber,
determined for the cases in which $\bar{\varepsilon}<0$, $\bar{\varepsilon}=0$, and
$\bar{\varepsilon}>0$. These calculations were performed using both the transfer-matrix method and
COMSOL simulations, an excellent agreement between the two approaches being observed. Similarly to
the case of topological surface modes existing at the interface between two planar plasmonic
superlattices with opposite signs of the averaged permittivity, these calculations reveal that, if
the sign of the averaged permittivity of the plasmonic Bragg fiber is negative, localized surface
modes always exist at the interface between the plasmonic Bragg fiber and the outermost uniform
dielectric medium, as shown in Fig. 2(a). On the other hand, no localized modes appear at the
interface between the outermost uniform dielectric medium and the plasmonic Bragg fiber with
$\bar{\varepsilon}=0$ and $\bar{\varepsilon}>0$, as shown in Figs. 2(b) and 2(c), respectively.

The condition for the existence of a localized surface state in the plasmonic Bragg fiber can be
related to the formation of surface plasmon polaritons (SPPs) at the interface between two
homogeneous and isotropic media. It is well know that SPPs exists only if permittivities of the
two media have opposite signs. As is illustrated in Figs. 3(a) through 3(d), indeed, the field
profile of the surface modes resembles that of SPPs, with an additional feature represented by the
field oscillations inside the plasmonic Bragg fiber. The similarity between the conventional SPPs
and surface modes in plasmonic Bragg fibers is also illustrated in Fig. 3(e), where we present the
dependence of the propagation constant of the fundamental ($m=0$) and  higher-order ($m=1,2$)
surface modes on the size of the plasmonic Bragg fiber. We note that, with the increase of the
number of dielectric-metall rings, the propagation constant of the fundamental and the
higher-order modes of the plasmonic Bragg fiber approach asymptotically that of conventional SPPs
formed at the planar interface between a semi-infinite metal and a semi-infinite dielectric
medium, namely, ${k_{\text{spp}}}={k_{0} }\sqrt{{\varepsilon _{M}}{\varepsilon _{D}}/({\varepsilon
_{M}}+{\varepsilon _{D}})}$. Although the propagation constants of the fundamental and
higher-order modes deviate from the dispersion curve of $k_{\text{spp}}$ when the number of
dielectric-metal rings decreases, localized surface modes exist for any number of rings. Moreover,
as shown in Fig. 3(f), when the plasmonic Bragg fiber with $\bar{\varepsilon}<0$ is replaced by a
homogeneous metallic cylinder with the same radius, the radial dependence of the propagation
constant of the surface modes is similar to that of the plasmonic Bragg fiber, as per Fig. 3(e).
We mention here that, the imaginary part of the propagation constants of these
surface modes is extremely small [$\mathfrak{Lm}(k_z)\sim 0.003$ for all the modes shown in
Fig.~3(e)], which defines an essentially long propagation length
($L_{\text{prop}}\sim\SI{58.8}{\micro\meter}$).

Direct numerical simulations of optical beams propagating in the plasmonic Bragg fibers
investigated in this work, performed by numerically solving the 3D Maxwell equations governing the
beam dynamics, corroborate the conclusions of the above analysis. Thus, Fig. 4 shows how an input
TM-polarized Gaussian beam evolves in the plasmonic Bragg fiber when the operational wavelength is
varied. The plasmonic Bragg fiber is composed of $n=30$ dielectric-metal rings, and its average
permittivity is tuned by varying the operational wavelength. As expected, when $\bar{\varepsilon}$
of the plasmonic Bragg fiber is negative, a localized mode quickly forms at the interface between
the fiber and the cladding, whereas the additional energy of the input wave diffracts off as
radiation waves, see Fig. 4(b). By contrast, the input optical beam strongly diffracts without any
signature of the formation of a surface mode, in the case when the plasmonic Bragg fiber has
$\bar{\varepsilon}=0$ or $\bar{\varepsilon}>0$, as shown in Figs. 4(c) and 4(d), respectively.
\begin{figure}[th]
\centering\includegraphics[width=12cm]{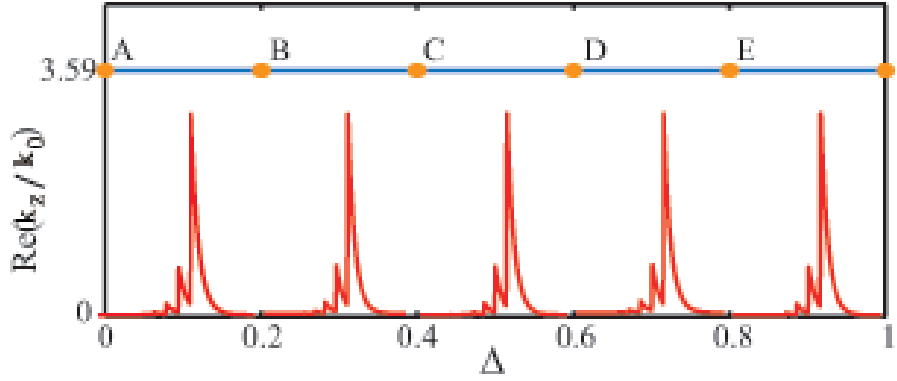}
\caption{The dependence of the eigenvalue (blue line) and electric field
amplitude of the fundamental surface mode on the disorder level, calculated
for the plasmonic Bragg fibers composed of $n=10$ dielectric-meta rings. The
operating wavelength is $\protect\lambda =\SI{2.2}{\micro\meter}$, thus the
average permittivity of the plasmonic Bragg fiber is negative, $\bar{\protect%
\varepsilon}<0$. The modal profiles are calculated for five disorder levels:
\SI{0}{\percent} (A), \SI{20}{\percent} (B), \SI{40}{\percent} (C),
\SI{60}{\percent} (D), \SI{80}{\percent} (E). All results are obtained by
averaging over $100$ disorder realizations.}
\end{figure}

As the existence of surface modes in the plasmonic Bragg fibers is
determined by the sign of the averaged permittivity, one expects that such
surface modes are extremely robust against structural disorder. This is
expected as a fully random structural perturbation preserves the average
values of the area of the transverse cross-section of constituent
cylindrical shells and thus does not modify the spatially averaged
permittivity or its sign. To test this conjecture, we introduce disorder
into the plasmonic Bragg fiber by assuming a random fluctuation of the area
of the transverse cross-section of the metallic components. Thus, the area
of the $n$th metallic layer is set to $\hat{S}_{M}^{n}=S_{M}^{n}+\delta _{n}$%
, where $\delta _{n}$ is a random value. We assume $\delta _{n}$ to be
uniformly distributed in the interval of $[-\delta ,\delta ]$, $0<\delta
<S_{M}^{1}$, hence the level of disorder can be characterized by the
parameter, $\Delta \equiv \delta /S_{M}^{1}$. We choose the operating
wavelength $\lambda =\SI{2.2}{\micro\meter}$ at which the unperturbed
plasmonic Bragg fiber has $\bar{\varepsilon}<0$. The eigenvalues and field
profiles of the fundamental surface modes determined for increasing disorder
strength, $\Delta $, are shown in Fig. 5, where the results are averaged
over 100 randomly-perturbed configurations.

As recently found in planar plasmonic lattices \cite{Deng}, the sign of the spatial
average of the permittivity determines the Zak phase of such superlattices, and the existence of
interfacial modes between a superlattice with $\bar{\varepsilon}<0$ and one with
$\bar{\varepsilon}>0$ can be naturally viewed as the edge modes occurring at the interface between
two topologically distinct structures. For plasmonic Bragg fibers considered here, although one
cannot rigorously define a Zak phase as in the case of 1D periodic structures, with the increase
of the number of metallic-dielectric ring pairs, the structure increasingly resembles a 1D
periodic structure as far as the outmost region is concerned. Therefore, one naturally expects
that similar topological modes occur in the regions where the average permittivity changes its
sign. From this topological viewpoint, the surface modes of plasmonic Bragg fibers with
$\bar{\varepsilon}<0$ are robust against structural disorder added to the photonic system. This 
envision is indeed confirmed by our analysis.  To be
more specific, as one can see in Fig. 5, the eigenvalue (propagation constant) of the surface
modes is unaffected by the structural disorder, and it is actually pinned to the value of the
conventional SPPs formed at the interface between a homogeneous metallic cylinder with the same
radius as the Bragg fiber and a dielectric cladding. Furthermore, the spatial profile of the
surface mode remains almost unchanged, even when the disorder strength increases to
\SI{80}{\percent} or even larger values. Therefore, the surface modes existing at the interface
between a plasmonic Bragg fiber with $\bar{\varepsilon}<0$ and uniform dielectric cladding may be
viewed as an extension of the topological modes that exist at the interface between two 1D
plasmonic superlattices with opposite sign of the average permittivity.

\section{Conclusion}

\label{sec:concl} Using both the mode analysis and direct beam propagation simulations, we have
studied the surface modes of the plasmonic Bragg fibers. Our analysis has revealed that the
existence of the surface modes in this setting is determined by the sign of the spatially averaged
electrical permittivity. As a consequence of this property, the surface states are robust against
addition of disorder to the system. The localized surface modes, which exist at the interface
between the Bragg-fiber's core with $\bar{\varepsilon}<0$ and the outermost uniform dielectric
medium, may be viewed as an extension of the topological modes in two 1D plasmonic superlattice
with opposite signs of the averaged permittivity in the corresponding periodic sublattices. 
Finally,  we mention that, it is known that the nonlinear optical effects are enhanced by the SPP effect, 
and the nonlinear change of the dielectric permittivity is even expected to change
the average permittivity from negative values to positive or vice versus, which can 
fundamentally change the topology of the metallic-dielectric Bragg 
structures  and consequently the  condition for the existence of the 
surface modes. Thus it  may be interesting to extend the present analysis to the 
light propagation in nonlinear Bragg fibers \cite{js}.

\section*{Funding}

European Research Council (ERC) (ERC-2014-CoG-648328); Guangdong Natural Science Foundation (Grant
Nos. 2015A030311018 and 2017A030313035); National Natural Science Foundation of China (NSFC) (61475101).

\end{document}